\documentclass[10pt,twocolumn,english,openbib]{article}
\usepackage{babel}
\usepackage{graphicx}
\usepackage{amsmath}
\usepackage{bm}
\usepackage{latexsym}
\usepackage{setspace}
\usepackage{amsfonts}

\usepackage{amssymb}   
\usepackage{amscd}
\usepackage{epsfig}
\usepackage{bbm}

\newtheorem{defi}{Definition}

\newtheorem{exempel}[defi]{Example}

\DeclareRobustCommand\openone{\leavevmode\hbox{\small1\normalsize\kern-.33em1}}%

\newcommand{\tr}{{\operatorname{Tr}}}

\newcommand{\bra}[1]{{\langle{#1}|}}
\newcommand{\ket}[1]{{|{#1}\rangle}}
\newcommand{\ketbra}[1]{{\ket{#1}\!\bra{#1}}}

\newlength{\blank}
\begin{document}

\twocolumn[
\begin{center}

\textbf{\Large How many copies are needed for state discrimination?}
\vspace{2mm} 

{\large 

Aram W. Harrow$^1$ and Andreas Winter$^2$

\textit{$^1$Department of Computer Science, University of Bristol,
        Bristol BS8 1UB, U.~K.}

\textit{$^2$Department of Mathematics, University of Bristol,
        Bristol BS8 1TW, U.~K.}


}
\end{center}
]

\thispagestyle{empty}

\noindent
{\bf The problem}
we are considering is motivated by the hidden subgroup
problem, for which the ``standard approach'' is to use the oracle to
produce the 
coset state $\rho_H = \frac{1}{|G|} \sum_{g\in G} \ketbra{gH}$, with 
$\ket{gH} = \frac{1}{\sqrt{|H|}} \sum_{h\in H} \ket{gh}$.
Determining $H < G$ then amounts to distinguishing the $\rho_H$, given
a small number of samples (disregarding complexity issues).

Abstractly, one is given a set of quantum states $\{ \rho_i : i=1,\ldots N \}$
on a $d$-dimensional Hilbert space ${\cal H}$, with the property that
the pairwise fidelities are bounded away from $1$:
\vspace{-1mm}\[
  \forall i\neq j \qquad
  F(\rho_i,\rho_j) := \| \sqrt{\rho_i}\sqrt{\rho_j} \|_1^2 \leq F < 1.
\vspace{-1mm}\]
The question is: how many copies of the unknown state $\rho_i$ does one need
to be able to distinguish them all with high reliability? In other words,
we would like to find, for $0 < \epsilon < 1$,
the minimal $n$ for which there exists a POVM 
$(M_i)_{i=1,\ldots,N}$ on ${\cal H}^{\otimes n}$ such that for all $i$,
$\tr (\rho_i^{\otimes n} M_i) \geq 1-\epsilon$.
Of course, this minimal $n$ will depend on the precise geometric position
of the states relative to each other, but useful bounds can be obtained
simply in terms of the number $N$ and the fidelity $F$.


\medskip\noindent
{\bf Upper bound.}
We invoke a result of Barnum and Knill~\cite{BarnumKnill} which says
that, assuming a probability distribution $(p_i)$ on the state set,
the average success probability is lower bounded as
\vspace{-1mm}\[\begin{split}
  P_{\rm succ} &:=   \sum_i p_i\tr(\rho_i^{\otimes n} M_i)                               \\
               &\geq 1 - \sum_{i\neq j} \sqrt{p_ip_j}                  
                                        \sqrt{F(\rho_i^{\otimes n},\rho_j^{\otimes n})}  
                \geq 1 - N\sqrt{F}^n,
\end{split}\vspace{-1mm}\]
which is $\geq 1-\epsilon$ if

\vspace{-1mm}\begin{equation}
  \label{eq:upper-bound}
  n \geq \frac{2}{-\log F}\left( \log N - \log\epsilon \right).
\vspace{-1mm}\end{equation}
In fact, this success probability is achieved by the ``square root''
or ``pretty good'' measurement~\cite{Holevo}, which, according
to~\cite{BarnumKnill}, has error probability not more than twice that of the
optimal measurement. So, for every distribution there exists
a POVM attaining success probability $\geq 1-\epsilon$. Conversely, for
fixed POVM one can try to find the worst probability distribution -- which
may be the point mass on the state with minimal
$\tr(\rho_i^{\otimes n} M_i)$. But looking
at the payoff function of this game, the success probability, we see
that it is bilinear in the strategies of the players, the probability
vector $(p_i)$ and the POVM $(M_i)$, and that furthermore the strategy
spaces of both players are convex. Hence, we can use the minimax
theorem~\cite{vonNeumannMorgenstern}:
\vspace{-1mm}\[
  \max_{(M_i)} \, \min_{(p_i)} \, P_{\rm succ}
           = \min_{(p_i)} \, \max_{(M_i)} \, P_{\rm succ} \geq 1-\epsilon,
\vspace{-1mm}\]
so there exists a POVM ${M_i}$ such that for all $i$,
$\tr (\rho_i^{\otimes n} M_i) \geq 1-\epsilon$.

\medskip\noindent
{\bf Lower bound.}
We quote from~\cite{Hayashi-etal}, the following lower bound (Theorem 1.4):
to distinguish the states $\rho_i$ with success probability $\geq \eta$,
\vspace{-1.5mm}\begin{equation}
  \label{eq:lower-bound}
  n \geq \frac{1}{\log(\lambda d)}\left( \log N + \log\eta \right)
\vspace{-1.5mm}\end{equation}
copies are necessary, where $\lambda := \max_i \| \rho_i \|$ is the largest
eigenvalue among the operators $\rho_i$.

\medskip\noindent
{\bf Applications and discussion.}
For constant $\eta$ and $\epsilon$, the upper and lower bounds
of eqs.~(\ref{eq:upper-bound}) and (\ref{eq:lower-bound}) are comparable,
provided $\lambda = O(1/d)$, which holds for many
important examples of the hidden subgroup problem.
Our upper bound
can be viewed as a generalisation and improvement of the results
in~\cite{Hayashi-etal} (Theorem 1.6),
which themselves improve on~\cite{Sen},
to the effect that $n=O(\log N)$ copies of a coset state are sufficient
to distinguish from among $N$ subgroups
(c.f. \cite{EHK99} which has $n=O(\log|G|)$ when specialising to the
hidden subgroup problem).  

Here, we get rid of assumptions on the group's structure (and indeed groups
at all), as well as a dimensional term in~\cite{Sen}.
Observe that by using the game theoretic trick (c.f.~\cite{KretschmannWerner})
we obtain a measurement with \emph{worst case} error $\epsilon$, unlike
previous approaches including~\cite{Hayashi-etal}.

\medskip\noindent
{\bf Acknowledgments.}
The authors thank Pranab Sen for providing the motivation for this work.
They acknowledge the hospitality of the Insitut Henri Poincare, Paris,
where the present work was done.

\noindent
Funding: U.K.~EPSRC (QIP IRC) and EU (QAP).

\vspace{-5mm}
\begin{spacing}{0.9}

\end{spacing}

\end{document}